\documentclass{aastex63}

\def\a{\"a}
\def\o{\"o}
\def\u{\"u}
\def\A{\"A}
\def\O{\"O}
\def\U{\"U}

\def\ang{\AA}

\def\gapprox{\lower.4ex\hbox{$\;\buildrel >\over{\scriptstyle\sim}\;$}}
\def\lapprox{\lower.4ex\hbox{$\;\buildrel <\over{\scriptstyle\sim}\;$}}
\def\ref#1{\par\noindent\hangindent1cm {#1}}

\begin{document}

\title{Global Energetics of Solar Flares. 
XI. Flare Magnitude Predictions of the GOES-Class}

\correspondingauthor{Markus J. Aschwanden}
\email{aschwanden@lmsal.com}

\author{Markus J. Aschwanden}

\affiliation{Solar and Stellar Astrophysics Laboratory (LMSAL),
 Palo Alto, CA 94304, USA}

\begin{abstract}
In this study we determine scaling relationships of 
observed solar flares that can be used to predict
 upper limits of the GOES-class magnitude of solar
 flares.  The flare prediction scheme is based on 
the scaling of the slowly-varying potential energy 
$E_p(t)$, which is extrapolated in time over an 
interval of $\Delta t \le$ 24 hrs. The observed 
scaling of the dissipated energy $E_{diss}$ scales 
with the potential field energy as $E_{diss} \propto 
E_p^{1.32}$. In addition, the observed scaling 
relationship of the flare volume, 
$V \propto E_{diss}^{1.17}$, the multi-thermal energy,
$E_{th} \propto V^{0.76}$, the flare emission measure 
$EM \propto E_{th}^{0.79}$, the EM-weighted temperature
$T_{w}$, and the GOES flux, $F_8(t) \propto E_p(t)^{0.92}$, 
allows us then to predict an upper limit of the
GOES-class flare magnitude in the extrapolated time window.
We find a good correlation (CCC$\approx 0.7$) between the 
observed and predicted GOES-class flare magnitudes (in 
172 X and M-class events). This is the first algorithm 
that employs observed scaling laws of physical flare
parameters to predict GOES flux upper limits, an important 
capability that complements previous flare prediction 
methods based on machine-learning algorithms
used in space weather forecasting.
\end{abstract}


\section{Introduction}

 Solar flare predictions, in particular for eruptive events
 that impact the heliospheric space weather, became an
 important area of research that may mitigate the health
 risk of astronauts in space, minimize power plant outages,
 and safeguard electronic equipment in the near-Earth space
 environment.
 An early warning system that could make short-term predictions
 within a time scale of 24 hours would be desirable. For the
 development of such a prediction tool we have to ask the
 question:  What are the most important physical parameters
 that facilitate the prediction of energies dissipated in
 solar flares and space weather events? Observables used
 for previous flare prediction methods include: the modified
 Zurich sunspot classification, the penumbra type of the
 largest sunspot, types of sunspot  distributions (Gallagher
 et al.~2002), high-gradient polarity-separation lines in
 line-of-sight magnetograms, the unsigned magnetic flux near
 the polarity-separation lines (Schrijver 2007), the effective
 connected magnetic field between unipolar magnetic areas in a
 flux-partitioned magnetogram (Georgoulis and Rust 2007), the
 length of the (neutral) polarity inversion line, the overall
 twist and shear of the non-potential magnetic field, the
 length segment over which the transverse magnetic field is
 strong (Falconer et al.~2003), the free magnetic energy 
(Falconer et al.~2011), the gradient-weighted inversion-line
 length (Mason and Hoeksema 2010; Cui et al.~2006), the
 unsigned current helicity, the total magnitude of the Lorenz
 force, the total photospheric magnetic free energy density,
 the total unsigned vertical current (Bobra and Couvidat 2015),
 the history of prior flaring (Falconer et al.~2012), the
 combination of UV brightenings, soft X-ray data, and vector
 magnetograph data (Nishizuka et al.~2017), synthesized
 photospheric and coronal data (Jonas et al.~2018), magnetic
 helicity injection in active regions (Tiwari et al.~2010),
 and measures of fractality, multi-fractality, and turbulence
 (Georgoulis 2012). This large arsenal of observables, which
 is still not capable to provide a fully satisfactory flare
 prediction method so far, indicates either that we have not
 yet identified the most relevant physical parameters, or that
 the measurement accuracy of the relevant physical parameters
 is insufficient. In both cases, machine-learning methods may
 not solve the problem (if the training data set does not
 contain the most relevant physical parameters), and thus we
 have to resort to more refined physical models. A benchmark
 test has been conducted during an inter-agency workshop on
 ``all-clear`` forecast, where the performance of a number of
 existing flare prediction algorithms was compared with common
 data sets from the {\sl Michelson Doppler Imager (MDI)}
 (Scherrer et al. 1995) onboard the {\sl Solar and
 Heliospheric Observatory (SOHO)} (Fleck et al. 1995),
 but it was found that no single method outperformed all
 others, similarly to the situation in climatological 
 forecasting (Barnes et al. 2016).

 In this study we develop a flare prediction method that is
 based on scaling relationships between magnetic field
 parameters (potential energy, non-potential energy, free
 energy, dissipated energy), geometric parameters (flare
 length scale, flare area, flare volume), temperature
 parameters (emission measure-weighted temperature), and
 energetic parameters (multi-thermal energy, emission measure,
 GOES flux). Most parameters have been previously measured
 in a series of papers on the global energetics of solar
 flares and {\sl coronal mass ejections (CME)}, in particular
 in Paper I (Aschwanden et al.~2014), Paper II (Aschwanden
 et al.~2015a), and Paper IX (Aschwanden 2019a). A fundamentally
 new aspect of this method developed here is the strategy to
 model the force-free coronal magnetic field with automatically
 traced coronal loop coordinates (e.g., Aschwanden 2016),
 which bypasses the problem of the non-forcefree photospheric
 magnetic field that is used in traditional magnetic field
 extrapolation methods (e.g., Wiegelmann et al. 2006). All
 the used observables and scaling relationships are based
 on data from the {\sl Helioseismic and Magnetic Imager (HMI)}
 (Scherrer et al.~2012) and the {\sl Atmospheric Imaging
 Assembly (AIA)} (Lemen et al.~2012) onboard the {\sl Solar
 Dynamics Observatory (SDO)} (Pesnell et al. 2012), and from
 the {\sl Geostationary Orbiting Earth Satellites (GOES)}
 (e.g., Lemen et al.~2004).

 The contents of this study include a summary of the most
 relevant observational results (Section 2.1-2.4), modeling results
 of flare predictions in terms of GOES-class magnitudes
 (Section 2.5), a discussion and comparisons with previous
 flare prediction methods (Section 3), and conclusions
 (Section 4).

\section{ Data Analysis and Results }

\subsection{Previous Observations and Parameter Measurements}

We use the same data set of solar flares presented in
Paper I (Aschwanden et al.~2014) and Paper II (Aschwanden 
et al.~2015a), which includes all M- and X-class flares 
observed with the SDO (Pesnell et al.~2012) during the first 
3.5 years of the mission (2010 June 1 to 2014 January 31). 
This selection provides 172 flare events with measurements 
of magnetic parameters (within a heliographic longitude range
of $[-45^\circ, +45^\circ]$, for which magnetic field modeling
can be faciliated without too severe foreshortening effects 
near the solar limb), and 391 flare events for measurements
of thermal parameters (for events at all longitudes $[-90^\circ, 
+90^\circ]$). 

We use the 45-s line-of-sight magnetograms from HMI/SDO and 
make use of all coronal EUV channels of AIA (Lemen et 
al.~2012) onboard the {\sl Solar Dynamics Observatory (SDO)} 
(in the six wavelengths 94, 131, 171, 193, 211, 335 \ang ), 
which are sensitive to strong iron lines in the temperature 
range of $T \approx 0.6-16$ MK.  The spatial resolution is 
$\approx1.6"$ (0.6" pixels) for AIA, and the pixel size of 
HMI is 0.5". 

The coronal magnetic field is modeled by using the 
line-of-sight magnetogram $B_z(x,y)$ from the 
{\sl Helioseismic and Magnetic Imager (HMI)} (Scherrer et 
al.~2012) and (automatically detected) projected loop 
coordinates $[x(s), y(s)]$ in each EUV wavelength of AIA.
A full 3-D magnetic field model ${\bf B}(x,y,z)$ is computed
for each time interval and flare with a cadence of 6 min,
where the total duration of a flare is defined
by the GOES flare start and end times. The size of
the computation box amounts to an area with a width and length
of 0.5 solar radii in the plane-of-sky, and an altitude range
of 0.2 solar radius. The total number of analyzed data
includes 2706 HMI images and 16,236 AIA images.

For the data analysis of this study, which is focused on 
scaling laws of magnetic energies dissipated during flares,
we extracted from the previous studies the following observables: 
the mean nonpotential magnetic field strength $B_{np}$,
the mean potential field strength $B_p$,  
the mean magnetic field component $B_{free}$
associated with the free energy $E_{free}$ (or azimuthal 
field component $B_\varphi$), the flare area $A$, and the total
magnetic energy $E_{diss}$ dissipated during the flare 
duration. The magnetic parameters $B_{np}$,
$B_\varphi$, $E_{free}$, $E_{diss}$ are all computed
with the {\sl Vertical-Current Approximation Nonlinear
Force-Free Field (VCA-NLFFF)} code, as described in Paper I
and Paper IX. This magnetic field extrapolation code 
essentially assumes vertical currents at flare locations 
that are associated with sub-photospheric magnetic field 
concentrations (e.g., sunspots and smaller magnetic features).
A major advantage of this code over
traditional NLFFF codes is the capability to measure the
current-driven twisting of coronal magnetic field lines, 
based on automated tracing of coronal loops in AIA images, 
which this way bypasses the non-force-freeness of the 
photospheric field. The spatial scale $L$ is measured from 
the area $A=L^2$ of the (time-accumulated) azimuthal 
magnetic field, i.e., $B_\varphi(x,y) \ge 100$ G, 
after correction of projection effects (Paper I).

The multi-thermal energy $E_{th}$ and mean electron
density $n_e$ is obtained from
a {\sl differential emission measure (DEM)} analysis, 
using the relationship of the total emission measure $EM$ 
with the electron density, i.e., $EM = n_e^2 V \approx 
n_e^2 L^3$ (Paper II). The quantities $B_{\varphi}, L, 
n_e$ are determined at the peak times of the flare 
emission measure, while the parameters $E_f$ and $E_{diss}$ 
are integrated over the flare duration $\tau$.

\subsection{ Observed Scaling Laws }

We quantify an observed solar flare with the observables 
$[E_p, A_{proj}, l, b, DEM(T)]$, where $E_p$ is the  potential
 magnetic energy (integrated over the volume of  a
 computational box that  encompasses a flaring active region),
 $A_{proj}$ is the projected 2-D area of the flaring region on
 the solar surface, $l$ and $b$ are the heliographic longitude
 and latitude, and $DEM(T)$ is a {\sl differential emission
 measure (DEM)} distribution. 

First we calculate the potential field energy $E_p$, which
can be obtained from a line-of-sight magnetogram $B_z(x,y)$
by means of a standard magnetic (potential) field extrapolation
in the 3-D volume $V$ of the computation box $[x,y,z]$ that 
encompasses a flaring active region,
\begin{equation}
	E_p = \int \left(
	 { B_p^2(x,y,z) \over 8 \pi } \right) \ dV \quad 
	 = \int \left( { B_p^2(x,y,z) \over 8 \pi }	
	 \right) \ dx \ dy \ dz \ .		
\end{equation}
Note that the potential field energy $E_p$ represents the
 minimum magnetic energy state, which is a lower limit of
 the nonpotential field (Priest 1975).

In Fig.~1a we show a linear regression fit to the free
 energies $E_{free}$ as a function of the potential field
 energies $E_p$, computed for 172 GOES X-class and M-class
 flares using the {\sl Vertical-current approximation
 non-linear force-free field (VCA-NLFFF)} code, according
to the latest version, as described
 in Paper IX (Fig.~8b therein). This linear regression fit
 yields the scaling relationship
\begin{equation}
	\left( { E_{free} \over 10^{30}\ {\rm erg} } \right)
	= 10^{-1.73} \ \left( { E_p \over 10^{30}\ {\rm erg} }
	 \right)^{1.26} \ .
\end{equation}
Similarly we show in Fig.~1b a linear regression fit to the
dissipated energies $E_{diss}$ as a function of the potential
field energies $E_p$, as described in Paper IX (Fig.~8c
therein). This linear regression fit yields the scaling
relationship
\begin{equation}
	\left( { E_{diss} \over 10^{30}\ {\rm erg} } \right)
	= 10^{-2.19} \ \left( { E_p \over 10^{30}\ {\rm erg} }
	 \right)^{1.32} \ .
\end{equation}
We see that these two relationships are similar, in particular
they have a similar power law slope of $p \approx 1.3$,
while the dissipated energy contains about half of the
free energy, i.e., $E_{diss} \approx E_{free}/2$.
Uncertainties of all linear regression fits are 
indicated in Figs.~1-2 by the mean (thick solid line)
and standard deviation of the slopes (thin solid lines).

In a next step we define the scaling relationships of
 geometric parameters, such as the flare length scale $L$,
 the flare area $A$, and the flare volume $V$. We can directly
 observe the projected flare area $A_{proj}$ only, and estimate
 the true (unprojected) area $A$ by dividing with the cosine
 of the longitude $l$ and latitude $b$ difference to disk
 center,
\begin{equation}
	A = { A_{proj} \over \cos{\sqrt{l^2 + b^2}} } \	.
\end{equation}
Note that the observed flare area $A_{proj}$ is measured
(in Papers I and II) from the sum of the pixel areas above some
threshold of the dissipated magnetic flare energy $E_{diss}$. 
The time dependence of the flare area is taken into account by
calculating the cumulative flare area, accumulated between
the flare start and end time (Papers I and II). After we have 
defined the flare area $A$, we can simply define a flare 
length scale $L$ by
\begin{equation}
	L = A^{1/2}  \ ,				
\end{equation}
and a flare volume $V$ by
\begin{equation}
	V = A^{3/2} \ .					
\end{equation}
As a next step we can investigate the relationship between
the dissipated energy $E_{diss}$ and the flare volume $V$,
which is shown in Fig.~1c (using the parameters determined
in Paper I and shown in Fig.~14c therein). The two
parameters exhibit a correlation that is fitted by the
relationship,
\begin{equation}
	\left( { V \over 10^{24}\ {\rm cm}^3 } \right)	
	= 10^{2.34} \ \left( { E_{diss} \over 10^{30}\ {\rm erg} }
	 \right)^{1.17} \ ,
\end{equation}
which is almost a linear function. 

Since both the flare volume $V$ and the (space-integrated)
multi-thermal energy $E_{th}$ have the same volume dependence,
we expect a nearly linear relationship. There is indeed a
strong correlation between these two parameters (CCC=0.91), 
as shown in Fig.~1d (or Fig.~5b in Paper II),
\begin{equation}
	\left( { E_{th} \over 10^{30}\ {\rm erg} } \right)
	= 10^{-1.54} \ \left( { V \over 10^{24}\ {\rm cm}^3 }
	 \right)^{0.76} \ .
\end{equation}
Another volume-dependent quantity is the emission measure 
$EM$, which is shown in Fig.~1e (or in Fig.~1f with reversed 
axes, as in Paper II, Fig.~5f therein),
\begin{equation}
	\left( { EM  \over 10^{40}\ {\rm cm}^{3} } \right)	
		= 10^{8.33} \ \left( { E_{th} \over 10^{30}\
	 {\rm erg} } \right)^{0.79} \ .
\end{equation}
Since the {differential emission measure (DEM)} is generally
a multi-temperature distribution, we define a weighted
temperature $T_w$, using the integral over the function
$DEM(T)$,
\begin{equation}
	T_w = {\int\ T \times DEM(T)\ dT \over \int\ DEM(T)\ dT} =
	  {\int\ T \times DEM(T)\ dT \over EM} \ . 
\end{equation}
The determination of the DEM distributions $DEM(T)$ using
AIA/SDO data is described in Paper II and references therein. 
The main feature of our DEM analysis is the proper treatement
of the multi-thermal distribution, rather than the commonly
used iso-thermal approximation. A benchmark test with
comparisons of 11 different DEM methods is described in
Aschwanden et al.~(2015b). In the study here we distinguish between
the (temperature-integrated) emission measure $EM_{AIA}$ 
calculated from the 6 coronal AIA channels, and the GOES 
emission measure $EM_{GOES}$, which is calculated from the 
iso-thermal channel ratio of the two GOES wavelength ranges
1-8 \ang\ and 0.5-4 \ang .

\subsection{ Scaling Law of GOES Flux } 

We can now use the two parameters $[EM, T_w]$ to predict
the GOES fluxes $F_4$ and $F_8$ in the wavelength ranges
$\lambda=0.5-4$ \ang\ and $1-8$ \ang. The original
calibration of the GOES fluxes is given in Thomas, Starr,
and Crannell (1985), and later updated by White, Schwartz,
and Thomas (2005). Software capabilities to extract GOES 
data have been implemented in the {\sl Interactive
Data Language (IDL)} {\sl Solar Software (SSW)}
by Freeland and Handy (1998).
We use the calibration of the most recent
GOES spacecraft (GOES-12) given in White et al.~(2005).
GOES data during the SDO era were provided by GOES-15
(launched on 2010 March 4) and by GOES-16 (launched on
2016 November 19). For a given flux ratio $R$,
\begin{equation}
	R = F_4 / F_8 \ ,				
\end{equation}
the GOES temperature $T_{GOES}$ is given by a third-order
 polynomial (with the coefficients measured from the most
 recent GOES satellite (GOES-12) given in that paper),
\begin{equation}
	T_{GOES} = 3.90 + 101.2\ R -266.4\ R^2 +390.2\ R^3 \ , 
\end{equation}
and the parameter $b_8(T)$ is given by a third-order polynomial also,			
\begin{equation}
	b_8(T) = {-12.31} + {3.75}\ T_{GOES}\ {-–0.1003}\ T_{GOES}^2
	 {+0.001195}\ T_{GOES}^3 \ ,
\end{equation}
yielding the GOES emission measure $EM_{GOES}$,		
\begin{equation}
	EM_{GOES} = 10^{55}\ F_8 / b_8(T) \ .		
\end{equation}
In our application we approximate the GOES emission measure
 $EM_{GOES}$ and temperature $T_{GOES}$  with the AIA-inferred
 emission measure $EM_{AIA}$ and emission measure-weighted
 temperature $T_w$. However, in order to allow for different
 temperature responses of the two instruments AIA and GOES, we define
 an empirical normalization factor $q_{AIA}$,
\begin{equation}
	q_{AIA} = { EM_{GOES} \over EM_{AIA} }  
	\approx 0.54 \ , 
\end{equation}  
which we determine by the normalization of the predicted
mean value of the GOES flux to the observed value,
i.e., $F_8^{pred}/F_8^{obs}=1.00$ (Fig.~3).
We can then directly calculate the
 calibration parameter $b_8$ (Eq.~14), and predict the GOES
 long-wavelength flux $F_8$ by inverting Eq.~(14),
\begin{equation}
	F_8 = {EM_{GOES}\ b_8 \over 10^{55}} =
 	      q_{AIA} \ EM_{AIA} \ {b_8 \ 10^{-55}} \ .
\end{equation}
The GOES flux is given in physical units of [W m$^{-2}$],
which scales with the well-known GOES flare classes as
$F_8=10^{-6}$ [W m$^{-2}$] for a C1.0-class flare,
$F_8=10^{-5}$ [W m$^{-2}$] for a M1.0-class flare, and
$F_8=10^{-4}$ [W m$^{-2}$] for a X1.0-class flare.

In Fig.~2 we show the resulting correlations of the calculated
GOES fluxes $F_8$ as a function of the emission measures
$EM$ (Fig.~2a), which represents the best-correlated parameter
(with a cross-correlation coefficient of CCC=0.99),
\begin{equation}
	\left( { F_8 \over {\rm W}\ {\rm m}^{-2} } \right)		
	= 10^{-14.19} \ \left( { EM \over 10^{40}\
	 {\rm cm}^{3} } \right)^{1.06} \ .
\end{equation}
There are also good correlations of the GOES flux $F_8$ with
the thermal energy $E_{th}$ (Fig.~2b), with a cross-correlation
coeffient of CCC=0.93, 
\begin{equation}
	\left( { F_8 \over {\rm W}\ {\rm m}^{-2} } \right)		
	= 10^{-5.41 } \ \left( { E_{th} \over 10^{30}\
	  {\rm erg} } \right)^{0.84} \ ,
\end{equation}
and with the thermal flare volume $V_{th}$, with a cross-correlation
coefficient of CCC=0.72,
\begin{equation}
	\left( { F_8 \over {\rm W}\ {\rm m}^{-2} } \right)		
	= 10^{-6.79} \ \left( { V_{th} \over 10^{24}\
	  {\rm cm}^3 } \right)^{0.67} \ .
\end{equation}

The most important test of our scaling law relationships is
the expected correlation between the observed GOES fluxes
$F_8^{obs}$ and the predicted values $F_8^{pred}$.
Remember that the GOES fluxes $F_8$ were calculated from the
emission measures $EM$ and the emission measure-weighted
temperatures $T_w$ (Eqs.~10-15), which are obtained from
the scaling laws of the emission measure $EM(E_{th})$ (Eq.~9), 
the (multi-)thermal energy $E_{th}(V)$ (Eq.~8), the
flare volume $V(E_{diss})$ (Eq.~7), the dissipated flare
energy $E_{diss}(E_p)$ (Eq.~3), and the potential field energy
$E_p(B_p)$ (Eq.~1). Thus the final prediction of the GOES
fluxes depends ultimately on the potential field
${\bf B}_p$ only. If our inferred scaling laws are correct,
we would expect a close correlation between the observed
$(F_8^{obs}$) and the predicted GOES fluxes ($F_8^{pred})$.
The analytical derivation of this GOES scaling law is given
in Appendix A.
We show a cross-correlation plot of these two quantities
in Fig.~(2d), which reveals a reasonably good cross-correlation
coefficient of CCC=0.71,
\begin{equation}
	\left( { F_8^{pred} \over {\rm W}\ {\rm m}^{-2} } \right)		
	= 10^{ 0.80} \ \left( { F_8^{obs} \over {\rm W}\ {\rm m}^{-2}\
	  } \right)^{1.18} \ .
\end{equation}
This test establishes our claim
that we can predict the GOES class of a flare, based on the
potential field alone. Since the temporal evolution of the
potential field is generally relatively slow, it provides
a robust prediction for the flare magnitude (i.e., GOES
class). 

In Fig.~3 we show the ratios $q_{acc}$ of the predicted 
$F_8^{pred}$ to the observed GOES values $F_8^{obs}$, 
which exhibit a Gaussian-like distribution with a mean and
standard deviation of $q_{acc}=1.8\pm0.6$, for the
172 flare events investigated in this study. Normalizing
the AIA-inferred multi-thermal emission measure $EM_{AIA}$
to the GOES-inferred isothermal emission measure $EM_{GOES}$,
we find a normalization factor of 
$q_{AIA}=EM_{GOES}/EM_{AIA}=0.54$, 
which has a mean and standard deviation of
$q_{AIA}=1.00 \pm 0.40$ (Fig.~3) and implies that
the accuracy of the GOES scaling laws is $\approx \pm 40\%$.

\subsection{ Simplified GOES Class Estimates }

Our goal is the prediction of the GOES (1-8 \ang ) flux $F_8$ for 
each flare, based on information on the potential field energy $E_p$. 
A direct scaling relationship between these two parameters
$F_8$ and $E_p$ can be obtained by combining all Eqs.~(1-15), 
\begin{equation}
	F_8 = q_{AIA} \ b_8(T_w) \ 10^{-8}\
	\left( {E_p \over 10^{30}\ {\rm erg}} \right)^{0.92} \quad 
	{\rm{[W\ m}^{-2}]} \ ,  
\end{equation} 
where the detailed analytical derivation is given in Appendix A.
Since the emission measure-weighted temperature in our sample
of 172 X and M-class flare events is relatively narrow, 
varying only by a factor of $\approx 20\%$,
\begin{equation}
	T_w \approx 25.5 \pm 5.6 \ {\rm [MK]} \ ,
\end{equation}
we obtain a mean calibration factor $b_8(T_w)$ of,
\begin{equation}
	b_8(T_w) \approx 38 \ ,
\end{equation}
which together with the normalization factor $q_{AIA}=0.54$
yields then the simple scaling law.
\begin{equation}
	F_8 \approx 2.05 \times 10^{-7} \ 
	\left( {E_p \over 10^{30}\ {\rm erg}} \right)^{0.92} 
	\quad [{\rm W\ m}^{-2}] \ .
\end{equation}
This simplified scaling law allows quick estimates of the GOES
class: Total (potential field) energies of $E_p \approx 5, 50$, and
500 $\times 10^{30}$ erg are required for flare magnitudes of
GOES class C1.0, M1.0, and X1.0. We will use the
simplified scaling law of Eq.~(24) for flare forecasting
in the following.  

\subsection{ Daily Forecasting of GOES Flare Magnitude }

In this section we generate a new data set that produces a
time series of magnetic potential field energies $E_p(t)$,
using the {\sl Vertical-Current Approximation Non-Linear 
Force-Free Field (VCA-NLFFF)} code. Detailed descriptions 
of the magnetic field code are given in Aschwanden et 
al.~(2014b, 2016). We use observations of active regions 
from HMI/SDO and AIA/SDO during 
the month of 2011 February 1-28, which we analyze with a
cadence of $\Delta t=24$ hours. The Sun is very active
during this month, producing one X2.2 GOES class flare
and 13 M-class flares (Table 1). We use the NOAA flare
catalog ({\sl hesperia.gsfc.nasa.gov}) to identify the 
heliographic coordinates of the active regions. There
are one, none, or multiple active regions present every day, 
for which we have to calculate magnetic field extrapolations
separately, and have to add them up in order to gather all
significant magnetic energies on the visible solar disk.
For the month of February 2011 we find a total of 23 days
(out of the 28 days) with active regions present, 
encompassing a total of 53 active regions, according to
the NOAA data catalog.   

Our main goal is to 
test whether our anticipated algorithm is able to predict
upper limits of the GOES 1-8 \ang\ flux $F_8(t)$ based on 
the potential magnetic energies only (i.e., using the
scaling law $F_8(t) \propto E_p(t)^{0.92}$ (Eq.~24). 
First we show the obtained time profiles of the potential
energy $E_p(t)$ (Fig.~4a), the nonpotential energy $E_{np}(t)$
(Fig.~4b),
and the free energy $E_{free}(t)$ (Fig.~4c). For comparison
we show also the results of an earlier version of the
VCA-NLFFF code (noisy time profiles in Figs.~4a and 4b)
for the time segment of 2011 Feburary 12-17 
(Aschwanden et al.~2016).
The time evolution of the potential and nonpotential 
energy is in good agreement between the old and new version
of the VCA-NLFFF code (Figs.~4a and 4b), 
while the free energy exhibits slight differences, caused by the
fact that the free energy is a small difference of two
large quantities, i.e., $E_{free}=E_{np}-E_p$, and thus
has a larger relative uncertainty.    

We show the GOES 1-8 \ang\ flux time profile is Fig.~4d
(grey area), which contains 14 M and X-class flare events 
during the investigated time window of 28 days. 
First we consider the ``now-cast'' case, where
the predicted time is identical to the observed time.
We take the potential field energy $E_p(t)$ and
calculate the upper limit of the GOES flux $F_8(t)$
with the scaling law given in Eq.~(24), i.e., 
$F_8(t) \propto E_p(t)^{0.92}$ and overplot it on the actually
observed GOES 1-8 \ang\ flux in Fig.~4d (grey area), 
which demonstrates that the so derived GOES fluxes 
indeed represent reliable upper limits as a function
of time. The uncertainty of the empirical scaling law 
amounts to $q_{acc}=F_8^{pred}/F_8^{obs}=1.0\pm0.4$ 
(Fig.~3). This means that we expect
a probability of 67\% for one standard deviation 
($\sigma=\pm 0.4$). 

Next we consider the ``flare prediction'' case 
(of the upper limit of the GOES flux $F_8^{pred}(t)$), 
for a time interval of $\Delta t=24$ hours ahead of the
observing time, which is shown in Fig.~4e (thick solid line).
The predicted value $E_p(t)$ is computed by linear 
extrapolation from the past two time intervals,
i.e., $E_p(t)=2*E_p(t-\Delta t)-E_p(t-2\Delta t)$,
and apply then the empirical scaling law $F_8(t) 
\propto E_p(t)^{0.92}$. To avoid unphysical negative
values in the extrapolated potential energy $E_p(t)$,
we extrapolate the logarithmic values. 
  
The four largest flares (see Table 1) occur
on 2011 February 13, 15, 18, and 24, with GOES fluxes of
M6.6, M2.2, M6.6, and M3.5 (indicated in Fig.~4e),
for which we predict upper limits of
M7.3, X1.4, M7.5, and M8.3, corresponding to flux ratios
of $q_8=F_8^{obs}/F_8^{pred}=$0.9, 1.5, 0.9, and 0.4,
which is approximately consistent with our expectation 
of $q_{acc} = 1.0 \pm 0.4$ (Fig.~3). 
Both the predicted upper limit of the GOES flux (Fig.~4e,
thick curve) and the limit of the now-cast case (Fig.~4e,
dashed curve), are shown together in Fig.~(4e), so that the 
uncertainty due to the temporal extrapolation can be seen. 
The comparison corroborates our assumption that the 
potential energy $E_p(t)$ is relatively slowly-varying on
time scales of a day.

\section{ Discussion }

\subsection{ Physical Parameters Relevant for Flare Prediction}

A magnetic potential field represents the minimum energy
 state (Priest 1975), from which no energy can be dissipated
 to produce a solar flare. Only the free energy
 $E_{free}=E_{np}-E_p$, which is the difference between the
 non-potential field and the potential field energy, can be
 dissipated in flares, which constrains the choice of valid
 physical parameters in flare models. A selection of 25
 physical parameters that can be measured in active regions
 has been presented by Bobra and Couvidat (2015), and a subset
 of 18 parameters thereof in Bobra and Ilionidis (2016),
 drawn from the {\sl Space-weather HMI Active Region Patches
 (SHARP)} project. For flare predictions, however, only those
 parameters are relevant that are directly sensitive to the
 free energy $E_{free} > 0$, or to the underlying electric
 current density $\nabla \times {\bf B} = 4\pi {\bf j}
 = \alpha({\bf r}) {\bf B}$.  In the study of Bobra and
 Couvidat (2015) we identify 17 relevant physical parameters,
 which quantify the free energy ($\rho_{tot}$, $\overline{\rho}$),
 the vertical currents ($J_{z,total}, J_{z,sum},
 \overline{J_z}$), the helical twist angle ($\alpha_{total}$),
 the shear angle $(\overline{\Gamma })$, the helicity
 $(H_{z,total}, H_{abs}, \overline{H_c})$,
 and the Lorentz force   
 ($F, F_z, F_y, F_x, \delta F_x, \delta F_y,v \delta F_z$). 
 The Lorentz force depends on the
 free energy component $B_{free}$ only, i.e.,
 ${\bf F} \propto {\bf j} \times {\bf B}_{np} =
 {\bf j} \times {\bf B}_p + {\bf j} \times {\bf B}_{free} =
 {\bf j} \times {\bf B}_{free}$, since there is no current
 in the potential field. 
 The other forthcoming
 parameters are sensitive to the nonpotential magnetic field
 strength ${\bf B}_{np}=[B_x, B_y, B_z]$, rather than to the
 magnetic field component associated with the free energy,
 i.e., ${\bf B}_{free}={\bf B}_{np}-{\bf B}_p$.  
 The remaining parameters, such as the 
 the magnetic flux ($\Phi, \Phi_R$), the flare area ($A, A_{45}$),
 the field divergence $(\overline{\nabla B_{tot}},
 \overline{\nabla B_z}, {\nabla B_h}$), or the inclination angle
 $\overline{\gamma}$, are thus not directly relevant for flare predictions,
 although they may be useful in characterizing (non-flaring)
 active regions.  

\subsection{  Flare Prediction Algorithms }

Early flare prediction algorithms were based on empirical
 (morphological) parameters rather than physical models.
 For instance, Gallagher et al. (2002) developed a flare
 prediction system which estimates the probability for
 each active region to produce C-, M-, or X-class flares
 based on sunspot classification. 

With increasing sophistication, a trend from morphological
 parameters to physical parameters took place, mostly measured
 from line-of-sight magnetograms. For instance, a method of
 partitioning the magnetic flux in a magnetogram and
 deriving the magnetic connectivity has been used in
 forecasting of major flares (Barnes et al.~2005;
 Georgoulis and Rust 2007).  The  lower moments of the field
 gradients, the kurtosis of the vertical current density, the
 magnetic twist, the current helicity density, and the
 magnetic shear angle has been used in another series of
 studies (Leka and Barnes 2003, 2007; Barnes et al.~2007;
 Barnes and Leka 2008).  It was noted that parameters that
 depend on the size of a flare are better correlated with the
 flaring behavior than parameters that do not depend on the
 system size (Welsch et al.~2009), which is consistent with
 our definition of the behavior of the free energy, which is
 spatially integrated over the entire flaring active region
 and thus depends on the system size.

Ultimately we aim to find physical flare models that 
can predict the onset of a flare instability, if solar
flares are produced by a deterministic process. However,
many studies show that flaring appears to be a
stochastic process, which cannot predict individual flares,
rather than a deterministic process. 
Neverthelss, flare forecasting has been approached by
machine-learning algorithms (e.g., Bobra and Couvidat 2015;
Bobra and Ilionidis 2016), which optimize prediction skills,
regardless of whether the flare process is stochastic or
deterministic. The algorithm of Bobra and Couvidat (2015)
is based on binary classifiers, where an
active region belongs to a positive class if it produces one
or more flares within a given time interval, and conversely,
an active region belongs to a negative class if it does not
produce a flare in the same time interval. This is slightly
different from our method with one qualifier, 
where only an upper limit of
a flare magnitude is predicted within a time window in the
near future. Thus, we do not predict the actual time when
a flare occurs, but only an upper limit in case a flare
occurs. This relieves also the class-imbalance problem
(Bobra and Couvidat 2015). In our method we can identify two types
of uncertainties: (i) The error of the theoretical scaling
law (which can be measured from the difference of theoretical
and observational values at now-cast time $t=t_{now}$), and
 (ii) the time extrapolation error (which can be measured from
 the difference in the prediction time window,
 at $(t > t_{now})$.  From the 25 physical parameters tested
 in Bobra and Couvidat (2015), we identify 10 relevant
 physical parameters that are sensitive to the free energy,
 the vertical current, the helical twist angle, the shear
 angle, or helicity, in agreement with our physical model
 of the {\sl Vertical Current Approximation (VCA)} NLFFF code.
 The underlying physical model is essentially a force-free
 solution of helically twisted magnetic field lines, which
 are generated by vertical currents emanating above
 (sub-photospheric) magnetic flux concentrations. If this
 physical model is correct, we expect that these physical
 parameters have a relatively high prediction score, which
 is indeed largely the case. However, the observables in the
 study of Bobra and Couvidat (2015) are based on photospheric
 magnetograms, which is not force-free (Metcalf et al.~1996),
 while the magnetic parameters in this study here are derived
 with the VCA-NLFFF code by fitting of coronal loops, which
 are thought to be force-free (with a plasma-$\beta$-parameter
 $\beta \ll 1$), and thus are expected to provide a more
 realistic magnetic field model.

\subsection{ GOES Fluxes and Flare Magnitude }

The GOES 1-8 \ang\ flux is often used to characterize the
flare magnitude, mostly because of the convenience of the
permanent availability of GOES observations (since 1974).
It is therefore important to quantify how well the GOES flux
correlates with other flare magnitude indicators. 
In this study we include large flares (of X and M-class)
only, which are the most relevant events to derive upper limits
of the GOES flux, while smaller flares (of C and B class) are
neglected here, since we are not interested in representative
distribution functions, which would exhibit power-laws
(e.g., Aschwanden and Freeland 2012).
The GOES flux appears to correlate with the upper limit of the
absolute total flux near the strong-field, high-gradient
polarity inversion lines (Schrijver 2007; 2009).
A strong correlation has been established between the
GOES flux and the thermal energy (Reep et al.~2013, 2020),
or between the temperature as well as the emission measure 
of the thermal plasma and the GOES flux (Warmuth et al.~2016a;
2016b). Even for cool (small) flares of GOES class B5 to C2,
the emission measure was found to be correlated with the
GOES flux (Phillips and Feldman 1995). Correlations between
the GOES flux and the AIA/SDO or EUVI/STEREO flux have been
found for occulted flares (Nitta et al.~2013). However, using
large statistics of 50,000 solar flares observed during three
solar cycles, the exact values of GOES peak temperatures and
emission measures were found to depend on the background
subtraction method (Ryan et al.~2012), originally pointed
out by Bornman (1990). However, for large (M and X-class) 
flares, as used here, the background correction is negligible.

We have to mention the caveat that the calculation of the 
GOES emission measure
and temperature from an inversion of the two-channel GOES
fluxes (1-8 \ang\ and 0.5-4 \ang ), is based on the 
isothermal assumption (Thomas et al.~1985; White et al.~2005),
and thus is generally not compatible when comparing 
different instruments, such as AIA, GOES, and RHESSI
(Ryan et al.~2014). Here we find an empirical normalization
factor for the emission measure obtained from GOES and AIA
(i.e., $q_{AIA} = EM_{GOES}/EM_{AIA} \approx 0.54$ (Eq.~15). 

In summary, while all previously published studies agree
that the emission measure $EM$ is highly correlated with
the GOES flux $F_8$ (Eq.~17), we find that the underlying
cross-correlation coefficient has actually the highest
value (CCC=0.99) among all investigated correlations,
and thus both the emission measure as well as the GOES
class (flux) are equally good measures of the flare magnitude,
analogously to the Gutenberg-Richter scale (or magnitude)
of earthquakes. 

\section{ Conclusions }

Solar flare prediction is a highly desirable goal, as the host 
of hundreds of scientific publications over the last decade
reveals. However, the terms ``prediction`` or ``forecasting`` 
have different meanings to researchers. We can
distinguish between at least two different types of
predictions: (i) the {\sl deterministic} approach,
which aims to predict the occurrence time and size of 
individuial events, and (ii) the {\sl stochastic} approach,
which predicts statistical distributions only, without 
any temporal evolution of individual events. 
The deterministic approach is of course the more challenging
task, because it requires an accurate and realistic physical model,
while the stochastic approach just requires statistics of
large samples. Extreme events are often produced by nonlinear
physical mechanisms, which have the tendency to form ``fat tails`` 
in their statistical distributions in form of power-law or 
log-normal functions. A popular new type of methods is ``machine-learning``,
which is able to learn and improve predictions by optimizing
parameter combinations from an initial ``training data set``.
However, successful machine-learning can only be accomplished
if the most relevant phyiscial parameters are included in the
training data set, which is essentially not available as long
as we do not have knowledge of the correct physical mechanism.
In this study we explore empirical scaling laws that quantify 
realistic parameter distributions and significant correlations 
of physical parameters in solar flares, which can
be extrapolated into the near future and predict reliable
upper limits of flare magnitudes (in terms of GOES fluxes).
Here the main conclusions:

\begin{enumerate}
\item{The success of any flare prediction algorithm rests in
the selection of relevant physical parameters. Since the
(magnetic) potential field represents the minimum state of
energy, only the free energy $E_{free}=E_{np}-E_p$ can be
dissipated in flares, which requires knowledge of the three
magnetic field components ${\bf B}_{p}$, ${\bf B}_{np}$,
and ${\bf B}_{free}$. Scaling laws that do not take into
account the free energy explicitly may be unphysical.}

\item{Another inconsistency that appeared in the extrapolation
of both the potential or the non-potential magnetic field is
the neglect of the photospheric non-force-freeness, a common
assumption in traditional NLFFF codes. Here we bypass the
non-force-freeness of the photosphere by fitting coronal loops
(which are supposed to be force-free in the corona due to the
low plasma-$\beta$ parameter), using the vertical current
approximation (VCA-NLFFF) code. This may affect the magnetic
field parameters and their correlations in flare scaling laws.} 

\item{Two-parameter correlations (Fig.~1) constitute a subset 
of simplified or approximative scaling laws, which we explore in 
this study. The best correlations among physical flare parameters
are found for the free energy and potential field energy, 
$E_{free} \propto E_p^{1.26}$; the dissipated energy, 
$E_{diss} \propto E_p^{1.32}$; the magnetic flare volume,
$V \propto E_{diss}^{1.17}$; the (multi-)thermal energy,  
$E_{th} \propto V_{th}^{0.76}$; the emission measure,
$EM \propto E_{th}^{0.79}$, and the GOES flux,
$F_8 \propto E_p^{0.92}$. The fact that these parameters 
exhibit tight correlations makes them highly relevant
for flare scaling laws.}

\item{In the opposite, uncorrelated parameters do not reveal 
scaling laws. This is evident, for instance, from scatterplots 
of the electron density $n_e$ or the electron temperature $T_e$ 
with other flare parameters, if we restrict to M and X-class
flares only. However, both the electron density and electron
temperature are correlated with the GOES flux if one includes
C and B-class flares (Feldman et al 1996). Here we restrict
ourselves to large flares of X and M-class for the 
prediction of upper limits of GOES fluxes, where 
the emission measure-weighted
temperature ($T_w = 25 \pm 5$ Mm) has a standard deviation
of $\pm 20\%$ only. The electron density $n_e$ has a similar small
spread of less than an order of magnitude (e.g., Aschwanden 2020), 
while all volume-dependent flare parameters exhibit a much larger 
spread (some 2-3 order of magnitudes for X- and M-class flares),
and thus dominate the two-parameter correlations.} 

\item{Among the parameters that correlate best with the GOES 
flux (Fig.~2), we find the emission measure, $F_8 \propto EM^{1.06}$; 
the thermal energy, $F_8 \propto E_{th}^{0.84}$;
and the thermal flare volume, $F_8 \propto V_{th}^{0.66}$.
Comparing the observed GOES-class values ($F_8^{obs}$) with
the theoretically predicted values ($F_8^{pred}$) based on the
two-parameter scaling laws we find a satisfactory agreement of
$q_{acc}=F_8^{pred}/F_8^{obs}=1.00 \pm 0.40$, after normalization
of the fluxes. Thus, the accuracy of predicted GOES-class 
magnitudes is of order $\pm 40\%$ (Fig.~3).}

\item{Our flare prediction method yields upper limits of the GOES flux 
$F_8^{theo}(t)$, either instantaneously for now-casting (Fig.~4d),
or during time intervals of at least $\Delta t \le 24$ hrs (Fig.~4e)
for fore-casting. The chief assumption is the slow time variability
of the potential field energy $E_p(t)$ on time scales of $\le 24$ hrs,
which allows the temporal extrapolation of the scaling law 
$F_8(t) \propto E_p(t)^{0.92}$ (Fig.~4). We tested our fore-casting
method over a time interval of 28 days and validated our claim 
that this method yields reliable upper limits of the GOES flux.
Future applications to longer time intervals with larger flare
statistics will allow us also to verify the statistical probabilities
of the most extreme events, which exceed the predicted upper limits,
also known as ``Dragon-King`` events (Sornette 2009; Sornette
and Ouillon 2012; Aschwanden 2019b).} 

\item{We have to be aware that two-parameter correlations
represent flare scaling laws with a minimal number of parameters,
which can be refined for specific flare models (such as the
Sweet-Parker or Petschek model). For instance, a Petschek-type
flare model can be quantified with 8 physical
parameters for the dissipated flare energy,
$E_{MR} \propto B_{free}^2 B_{np} L n_e^{1/2} \lambda M_A \tau_{flare}$,
with $E_{MR}$ the dissipated energy, $B_{free}$ the magnetic field
component associated with the free energy,
$B_{np}$ the non-potential field strength, $L$ the flare length
scale, $n_e$ the electron density, $\lambda$ the electron
density scale height, $M_A$ the Mach number of the magnetic
reconnection outflow, and $\tau_{flare}$ the flare duration
(Aschwanden 2020).}  

\end{enumerate}

The aim of this study is the distillation of the most relevant
physical parameters that play a role in flare prediction methods.
Once we manage this task, deterministic as well as
stochastic flare prediction becomes a promising option,
at least over a time interval that exhibits coherent 
magnetic field evolution. Future work may test the reliability
of predicting upper limits for the GOES flux over the entire
10-year interval of the SDO mission. Furthermore, various
machine-learning methods applied to the most relevant physical
parameters may optimize realistic scaling laws.

\vskip1cm
Acknowledgements:
{\bf We thank the anonymous referee and the statistical editor for
helpful comments.} 
We acknowledge software support by Samuel Freeland, Greg Slater,
and Mark Noga. 
Part of the work was supported by NASA contract NNG 04EA00C of the
SDO/AIA instrument and the NASA STEREO mission under NRL contract
N00173-02-C-2035.

\section{ Appendix A : Derivation of Scaling Laws for GOES Flux }

The scaling laws between the magnetic potential field energy $E_p$,
the dissipated energy in flares $E_{diss}$, the thermal flare volume $V_{th}$,
the (multi-)thermal flare energy $E_{th}$, and the emission measure $EM$
are specified in Eqs.~(9, 8, 7, 3), 
$$
	\left( {EM \over 10^{40}\ {\rm cm}^3 } \right)
	= 10^{a_0} \left( {E_{th} \over 10^{30}\ {\rm erg} } \right)^{a_1} \ ,
	\eqno(A1)
$$
$$
	\left( {E_{th} \over 10^{30}\ {\rm erg} } \right)
	= 10^{b_0} \left( {V_{th} \over 10^{24}\ {\rm cm}^3 } \right)^{b_1} \ ,
	\eqno(A2)
$$
$$
	\left( {V_{th} \over 10^{24}\ {\rm cm}^3 } \right)
	= 10^{c_0} \left( {E_{diss} \over 10^{30}\ {\rm erg} }\right)^{c_1}\ ,
	\eqno(A3)
$$
$$
	\left( {E_{diss} \over 10^{30}\ {\rm erg} } \right)
	= 10^{d_0} \left( {E_{p} \over 10^{30}\ {\rm erg} }\right)^{d_1}\ ,
	\eqno(A4)
$$
with the coefficients 
$a_0=8.33$, $a_1=0.79$,
$b_0=-1.54$, $b_1=0.76$,
$c_0=2.34$, $c_1=1.17$,
$d_0=-2.19$, $d_1=1.32$,
as determined from the linear regression fits shown in Fig.~1.
The scaling law of the GOES flux $F_8$ as a function of the emission
measure $EM$ is according to Eq.~(16),
$$	
	F_8(EM) = f_0 \ EM \ , \qquad f_0= q_{AIA} b_8(T_w) 10^{-55} \ .
	\eqno(A5)
$$
The second relationship of the GOES flux $F_8(E_{th})$ as a function
of the thermal energy $E_{th}$ is obtained by inserting Eq.~(A1) into Eq.~(A5),
$$
	F_8(E_{th}) = f_0\ 10^{40+a_0}
	\left( {E_{th} \over 10^{30}\ {\rm erg}} \right)^{a_1} \ .
	\eqno(A6)
$$ 
The third relationship of the GOES flux $F_8(V_{th})$ as a function
of the thermal flare volume $V_{th}$ is obtained by inserting Eq.~(A2) into Eq.~(A6),
$$
	F_8(V_{th}) = f_0\ 10^{40+a_0+b_0 a_1}
	\left( {V_{th} \over 10^{24}\ {\rm erg}} \right)^{a_1 b_1} \ .
	\eqno(A7)
$$ 
The fourth relationship of the GOES flux $F_8(E_{diss})$ as a function
of the magnetic dissipated energy $E_{diss}$ is obtained by inserting Eq.~(A3)
into Eq.~(A7),
$$
	F_8(E_{diss}) = f_0\ 10^{40+a_0+b_0 a_1+c_0 a_1 b_1}
	\left( {E_{diss} \over 10^{30}\ {\rm erg}} \right)^{a_1 b_1 c_1} \ .
	\eqno(A8)
$$ 
The fifth relationship of the GOES flux $F_8(E_{p})$ as a function
of the magnetic potential energy $E_{p}$ is obtained by inserting Eq.~(A4)
into Eq.~(A8), 
$$
	F_8(E_{p}) = f_0\ 10^{40+a_0+b_0 a_1+c_0 a_1 b_1+d_0 a_1 b_1 c_1}
	\left( {E_p \over 10^{30}\ {\rm erg}} \right)^{a_1 b_1 c_1 d_1} \ .
	\eqno(A9)
$$ 
which yields to the numerical values given in Eqs.~(21) and (24). 
This scaling law stated in Eq.~(24) is most useful for flare prediction
in terms of the GOES flux $F_8$ as a function of the magnetic potential
field energy $E_p$.  

\clearpage

%%%%%%%%%%%%%%%%%%%%%%%%% REFERENCES %%%%%%%%%%%%%%%%%%%%%%%%

\section*{ References }

\def\ref#1{\par\noindent\hangindent1cm {#1}}

\ref{Aschwanden, M.J. and Freeland, S.L. 2012, A\&A 754, 112} 
\ref{Aschwanden, M.J., Xu, Y., and Jing, J. 2014a, ApJ 797:50 (Paper I)}
\ref{Aschwanden, M.J., Sun, X., and Liu, W. 2014b, ApJ 785:34}
\ref{Aschwanden, M.J., Boerner, P., Ryan, D., Caspi, A., McTiernan, J.M.,
	and Warren, H.P. 2015a, ApJ 802:53 (Paper II)}
\ref{Aschwanden, M.J., Boerner, P., Caspi, A., McTiernan, J.M.,
	Ryan, D., and Warren, H. 2015b, SoPh 290, 2733}
\ref{Aschwanden, M.J. 2016, ApJSS 224, 225}
\ref{Aschwanden, M.J. 2019a, ApJ 885:49 (Paper IX)}
\ref{Aschwanden, M.J. 2019b, ApJ 880:105}
\ref{Aschwanden, M.J. 2020, ApJ (in press), 
	{\sl Global Energetics of Solar Flares: Petschek Reconnection
	Rate and Alfv\'en Mach Number of Magnetic Reconnection Outflows}}
\ref{Barnes, G., Longcope, D.W., and Leka, K.D. 2005, ApJ 629, 561}
\ref{Barnes, G., Leka, K.D., Schumer, E.A., and Della-Rose, D.J. 2007, 
	Space Weather 5,9, CiteID S09002}
\ref{Barnes, G. and Leka, K.D. 2008, ApJ 688, L107}
\ref{Barnes, G., Leka, K.D., Schrijver, C.J., et al.  2016, ApJ 829, 89}
\ref{Bobra, M.G. and Couvidat, S. 2015, ApJ 798:135}
\ref{Bobra, M.G. and Ilionidis, S. 2016, ApJ 821:127}
\ref{Bornmann P.L. 1990, ApJ 356, 733} 
\ref{Cui, Y., Li, R., Zhang, L. et al. 2006, SoPh 237, 45}
\ref{Falconer, D.A., Moore, R.L., and Gary, G.A. 2003, JGR 108, A10, 1380}
\ref{Falconer, D.A., Abdulnasser, F., Khazanov, I. et al. 2011,
	 Space Weather 9/4, S04003}
\ref{Falconer, D.A., Moore, R.L., Barghouty, A.F. et al. 2012, ApJ 757, 32}
\ref{Feldman, U., Doschek, G.A., Behring, W.E., and Phillips, K.J.H.
	1996, ApJ 460, 1034}
\ref{Fleck, B., Domingo, V., and Poland, A. (eds.) 1995, SoPh 162}
\ref{Freeland, S.L. and Handy, B.N. 1998, SoPh 182, 497}
\ref{Gallagher, P.T., Moon, Y.J., and Wang, H. 2002, SoPh 209, 171}
\ref{Georgoulis, M.K. and Rust, D.M. 2007, ApJ 661, L109.}
\ref{Georgoulis, M.K. 2012, SoPh 276, 161} 
\ref{Jonas, E., Bobra, M., Shankar, V. et al. 2018, SoPh 293, 48}
\ref{Leka, K.D. and Barnes G. 2007, ApJ 656, 1173}
\ref{Leka, K.D. and Barnes, G, 2003, ApJ 595, 1277}
\ref{Lemen, J.R., Title, A.M., Akin, D.J., et al. 2012, SoPh 275, 17}
\ref{Lemen, J.R., Duncan, D., Edwards, C., et al. 2004, SPIE 5171, 65}
\ref{Metcalf, T.R., Litao, J., McClymont, A.N., and Canfield, R.C.,
	 1995, ApJ  439, 474}
\ref{Nishizuka, N., Sugiura, K., Kubo, Y., Den, M., Watari, S., and Ishii, M.	
          2017, ApJ 835, 156}
\ref{Nitta, N.V., Aschwanden, M.J., Boerner, P.F., Freeland, S.L.,
	Lemen, J.R., and Wuelser, J.P. 2013, SoPh 288, 241}
\ref{Pesnell, W.D., Thompson, B.J., and Chamberlin, P.C.  2012, SoPh 275, 3}
\ref{Phillips, K.J.H. and Feldman, U. 1995, A\&A 304, 563}
\ref{Priest, E.R. 1975, SoPh 43, 177}
\ref{Reep, J.W., Bradshaw, S.J., and McAteer, R.T.J. 2013,
	ApJ 778:76} 
\ref{Reep, J.W., Antolin, P., and Bradshaw, S.J.et al. 2020, ApJ 890, Id. 100}
\ref{Ryan, D.F., Gallather, P.T., Dennis, B.R., Tolbert, A.K.,
	Schwartz, R.A., and Young, C.A. 2012, ApJSS 202, id. 11}  
\ref{Ryan, D.F., O'Flannagaink A.M., Aschwanden, M.J.,
	and Gallagher P.T. 2014, SoPh 289, 2547}
\ref{Scherrer, P.H., Bogart, R.S., Bush, R.I. et al. 1995, SoPh 162, 129}
\ref{Scherrer P.H., Schou, J., Bush, R.J. 2012, SoPh 275, 207}
\ref{Schrijver, C.J., Sandman, A.W., Aschwanden, M.J.,
	and DeRosa, M.L. 2004, ApJ 615, 512} 
\ref{Schrijver, C.J. 2007, ApJ 655, L117} 
\ref{Schrijver, C.J. 2009, Adv. Space Res. 43, 739}
\ref{Sornette, D. 2009, J. Terraspace Science and Engeneering, 2, 1}
\ref{Sornette, D. and Ouillon, G. 2012, EPJST, 205, 1} 
\ref{Tiwari, S.K., Venkatakrishnan, P., and Gosein, S. et al. 2010, ApJ 721, 622.}
\ref{Thomas, R.J., Starr, R., and Crannell, C.J. 1985, SoPh 95, 323}
\ref{Warmuth, A. and Mann, G. 2016a, A\&A 588, A115}
\ref{Warmuth, A. and Mann, G. 2016b, A\&A 588, A116}
\ref{Welsch, B.T., Li, Y., Schuck, P.W., and Fisher, G.H. 2009, ApJ 705, 821}
\ref{Wheatland, M.S. 2005, Space Weather 3/7, S07003}
\ref{White, S.M., Thomas, R.J., and Schwartz, R.A. 2005, SoPh 227, 231}
\ref{Wiegelmann, T., Inhester, B., and Sakurai, T. 2006, SoPh 223, 215}
 
\clearpage

%%%%%%%%%%%%%%%%%%%%%%% TABLES %%%%%%%%%%%%%%%%%%%%%%%%%%%%%

\begin{table}[h] 
\caption{Observed and predicted upper limits of the GOES
1-8 \ang\ flux during the month of 2011 February 1-28,
containing a total of 14 X and M-class flares.}
\begin{tabular}{llllll}
\hline
Date Time       & Heliographic & NOAA & GOES     & Predicted    & Ratio \\
		& position &	  & flux         & upper limit  &       \\
 	        &          &	  & $F_8^{obs}$  & $F_8^{pred}$ & $F_8^{obs}/F_8^{pred}$ \\ 
                &          &      & [W m$^{-2}$] & [W m$^{-2}$] &       \\ 
\hline
\hline
2011-02-09 01:23:00 & N16W70 & 11153 & M1.9 & M1.0 &  1.9 \\
2011-02-13 17:28:00 & S21E04 & 11158 & M6.6 & M7.3 &  0.9 \\
2011-02-14 17:20:00 & S20W07 & 11158 & M2.2 & X1.1 &  0.2 \\
2011-02-15 01:44:00 & S21W12 & 11158 & X2.2 & X1.4 &  1.5 \\
2011-02-16 01:32:00 & S22W27 & 11158 & M1.0 & M7.7 &  0.1 \\
2011-02-16 07:35:00 & S23W30 & 11161 & M1.1 & M6.7 &  0.2 \\
2011-02-16 14:19:00 & S23W33 & 11158 & M1.6 & M5.8 &  0.3 \\
2011-02-18 09:55:00 & S21W55 & 11158 & M6.6 & M7.5 &  0.9 \\
2011-02-18 10:23:00 & N17E07 & 11162 & M1.0 & M7.4 &  0.1 \\
2011-02-18 12:59:00 & S20W70 & 11158 & M1.4 & M7.2 &  0.2 \\
2011-02-18 14:00:00 & N17E04 & 11162 & M1.0 & M7.1 &  0.1 \\
2011-02-18 20:56:00 & N15E00 & 11162 & M1.3 & M6.4 &  0.2 \\
2011-02-24 07:23:00 & N14E87 & 11163 & M3.5 & M8.3 &  0.4 \\
2011-02-28 12:38:00 & N22E35 & 11164 & M1.1 & M1.0 &  1.1 \\
\hline
\end{tabular}
\end{table}
\clearpage

%%%%%%%%%%%%%%%%%%%%%%% FIGURES %%%%%%%%%%%%%%%%%%%%%%%%%%%%%

\begin{figure}
\centerline{\includegraphics[width=1.0\textwidth]{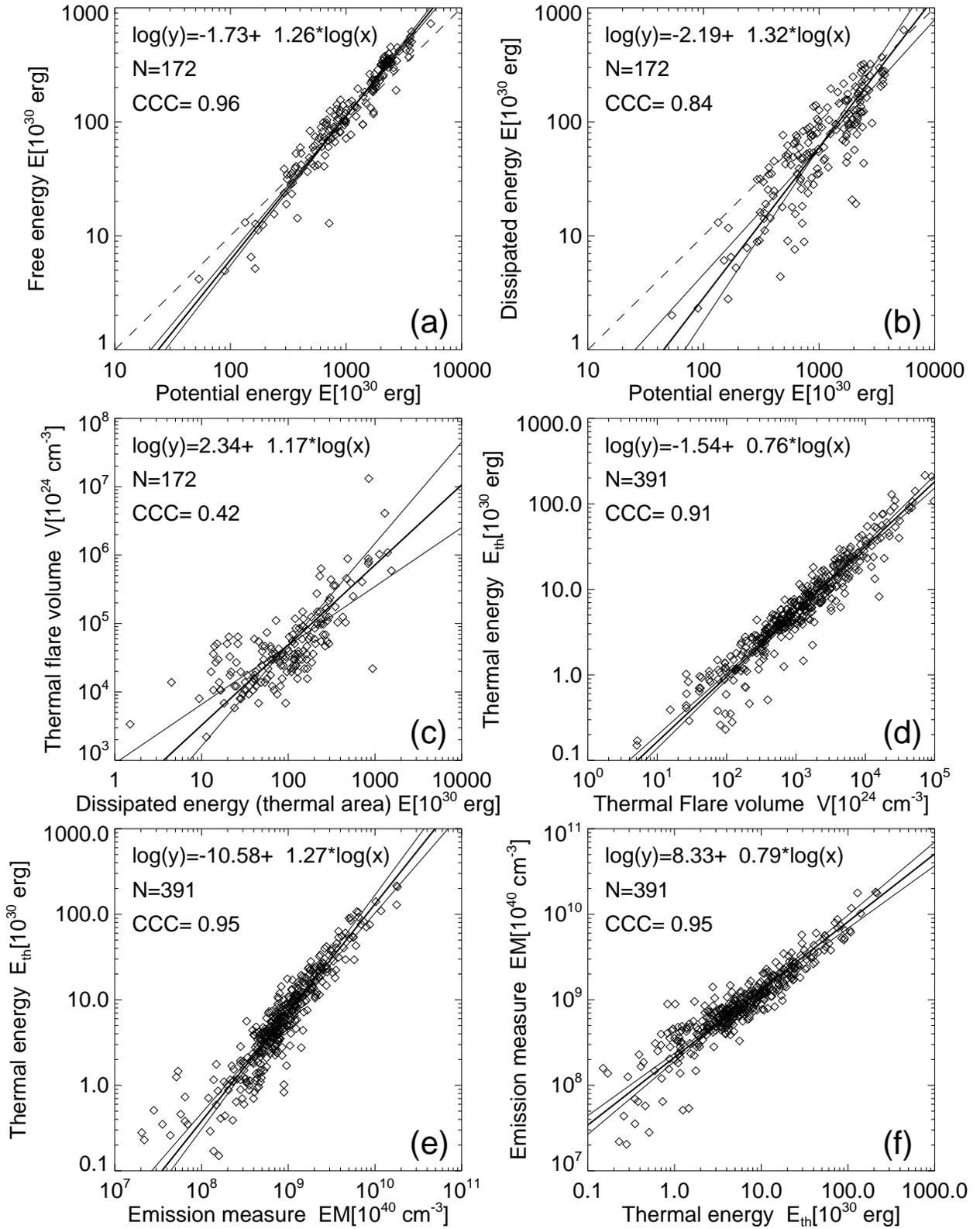}}
\caption {Correlations between the observed values (diamonds)
as a function of the potential energy are shown for the free
energy (a), the dissipated energy (b), the flare volume (c),
the thermal energy (d), and the emission measure (e,f). Linear 
regressions are shown for the logarithmic values, with the
mean (thick solid lines) and standard deviation (thin solid
lines), and equivalence (dashed lines) indicated.}
\end{figure}

\begin{figure}
\centerline{\includegraphics[width=1.0\textwidth]{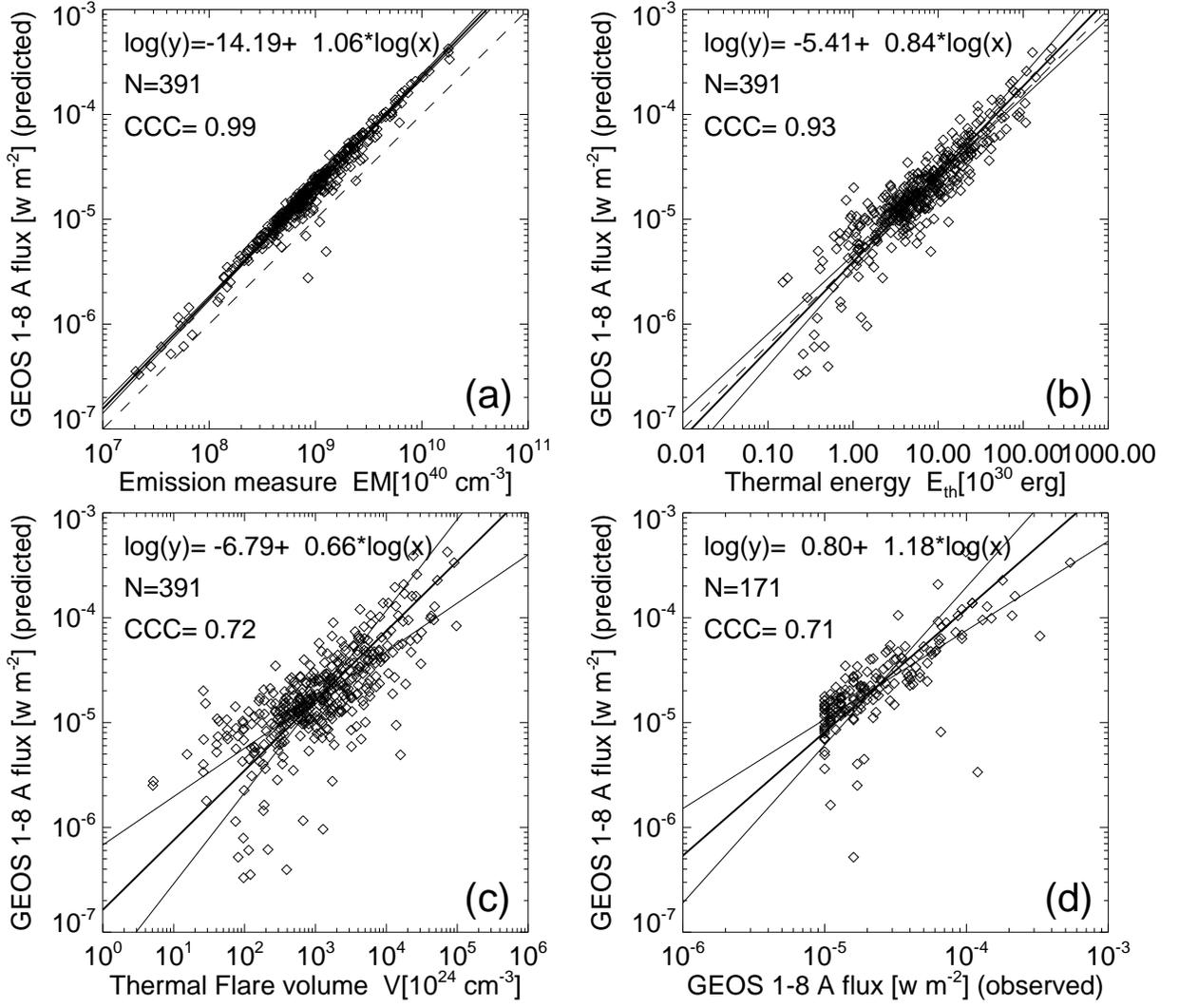}}
\caption{Correlations between the calculated GOES class and
the emission measure $EM$ (a), the (multi-)thermal energy
$E_{th}$ (b), the flare volume $V$, and the observed GOES
class (d). Linear regressions are shown for the logarithmic 
values, with the means (thick solid lines) and standard 
deviations (thin solid lines), and equivalence (dashed lines)
indicated.}
\end{figure}

\begin{figure}
\centerline{\includegraphics[width=1.0\textwidth]{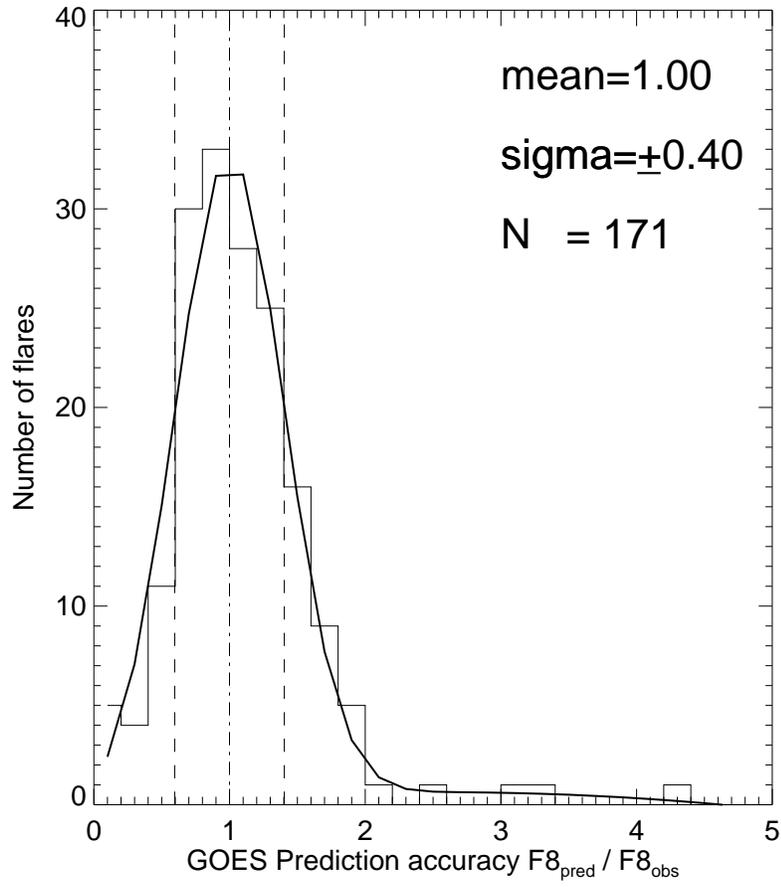}}
\caption{Histogram of the GOES class prediction accuracy, 
which is specified by the ratio of $q_{acc}=F8_{pred}/F8_{obs}$ 
with a mean and standard deviation of $q_{acc}=1.0\pm0.4$,
shown as Gaussian fit.} 
\end{figure} 

\begin{figure}
\centerline{\includegraphics[width=0.9\textwidth]{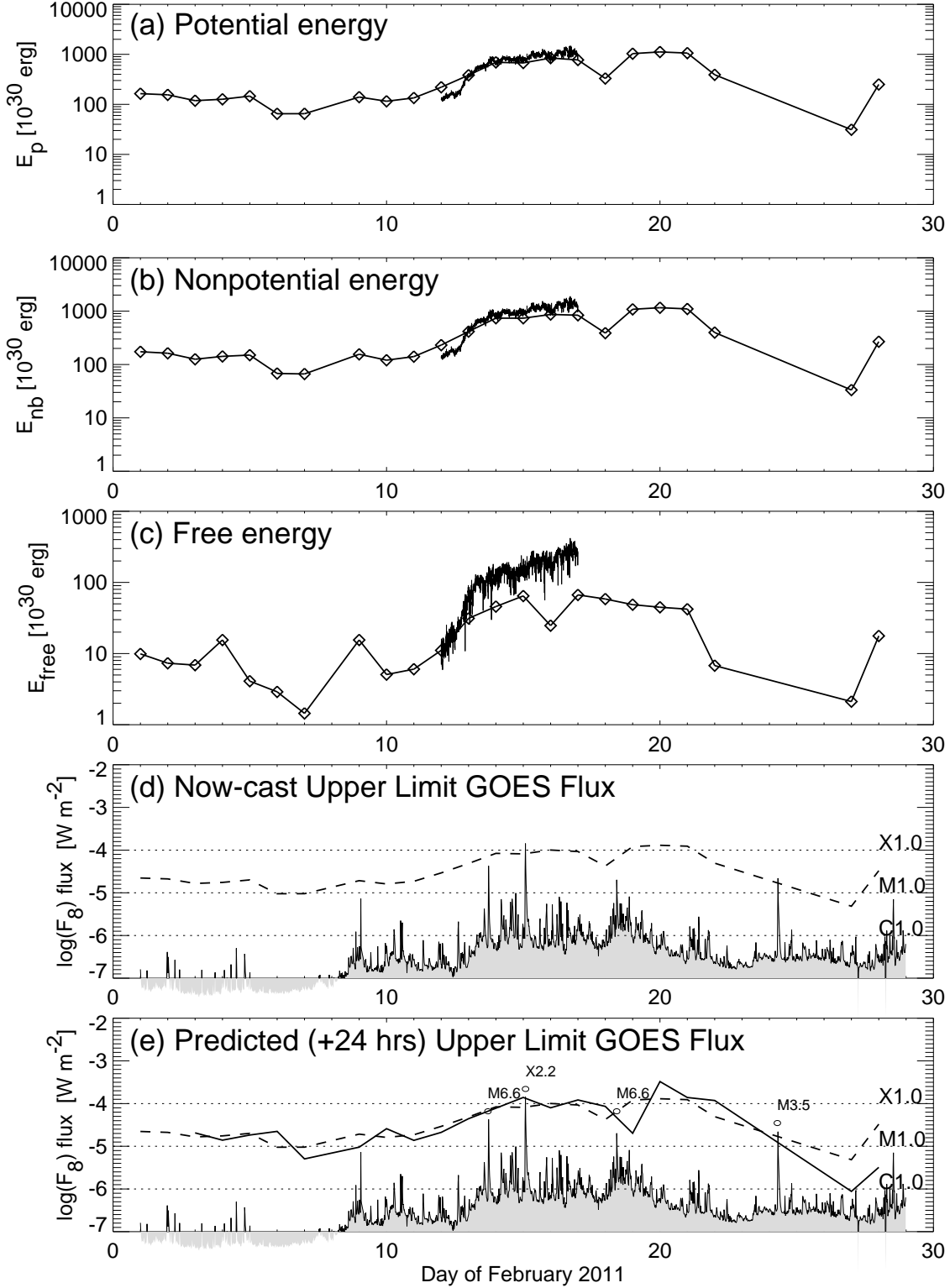}}
\caption{Time evolution of the potential-field energy $E_p$ (a),
the non-potential energy $E_{np}$ (b), the free energy $E_{free}
=E_{np}-E_p$ (c) during the month of February 2011, with a daily
cadence. The light curves (diamonds) are new calculations
with the latest version of the VCA-NLFFF code, while the noisy 
curves with cadence of 6 minutes are taken from an older
version (Aschwanden 2016).
Panel (d) shows the GOES light curve (shaded with gray),
with a cadence of 0.3 hrs, and the
now-cast upper limit of the GOES flux (dashed curve) calculated 
from the scaling relationship $F_8(t) \propto E_p(t)^{0.92}$. 
Panel (e) shows the predicted upper limit of the GOES 1-8 \ang\
flux, extrapolated for a time window of $+24$ hours
(solid curve), along with the now-cast predictions (dashed curve).
Four observed flare events of GOES class $>M3.0$ are marked with
circles, which all are close to the predicted upper limit of the
GOES flux.}
\end{figure} 

\end{document}